\documentclass[12pt,a4paper]{article}
	
\usepackage{a4wide}
\usepackage{amsmath}
\usepackage{amssymb}
\usepackage{amsfonts}
\usepackage{epsfig}
\usepackage{exscale}
\usepackage{float}
\usepackage{bbm}
\usepackage[numbers,sort&compress]{natbib}

\makeatletter
\@addtoreset{equation}{section}
\makeatother

\title{
A Solution of the Relativistic Schr\"odinger\\ Equation for the $\delta$-Function Potential \\in 1-dimension
with Cutoff Regularization}

\author{M.\ H.\ Al-Hashimi$^{a,b}$  and  Abouzeid M.\ Shalaby$^{a,c}$
\footnote{Contact information: M.\ H.\ Al-Hashimi: hashimi@itp.unibe.ch,
+41 31 631 8878; A.\ Shalaby, amshalab@qu.edu.qa, +974 4403 4630.}
\\ \\
$^a$ Department of Mathematics, Statistics, and Physics \\
Qatar University, Al Tarfa, Doha 2713, Qatar \\ \\
$^b$ Albert Einstein Center for Fundamental Physics \\
Institute for Theoretical Physics, Bern University \\
Sidlerstrasse 5, CH-3012 Bern, Switzerland \\ \\
$^c$ Physics Department, Faculty of Science \\
Mansoura University, Egypt \\ \\}

\begin{document}

\maketitle

\vspace{-1cm}

\begin{abstract} \normalsize
We study the  solution of the relativistic Schr\"odinger equation for a point particle in 1-d under $\delta$-function potential by using cutoff regularization. We show that the problem is renormalizable, and the results are exactly the same as the ones obtained using dimensional regularization.
\end{abstract}
\section{Introduction}
Contact interactions have been studied thoroughly in non-relativistic quantum mechanics. Different studies have illustrated some of the analogies between quantum mechanics and quantum field theory
\cite{Cal88,Tho79,Beg85,Hag90,Jac91,Fer91,Gos91,Mea91,Man93,Phi98}. Nevertheless, the analogies are still to a large extent ambiguous. In our opinion, studying contact interactions relativistically give a better chance for understanding the relation between quantum mechanics and quantum field theory, especially quantum field theory is basically a theory for describing the interaction of relativistic particles.
The problem of relativistic $\delta$-function potential  has been studied in the mathematical
literature in the  context of the  theory of self-adjoint extensions of
pseudo-differential operators \cite{Alb97}. Lately, we studied the problem by directly solving the relativistic version of the Schr\"odinger equation of the Hamiltonian $H = \sqrt{p^2 + m^2}+\lambda\delta(x)$ in 1-d \cite{Our}. Unlike the textbook case of the  non-relativistic $\delta$-function potential, the relativistic problem  gives rise to ultra-violet
divergences that required regularization  and renormalization. Previously, we used dimensional regularization to show that the system has remarkable features. For example, the relatively simple system shares many features with some complex quantum field theories, like asymptotic freedom, dimensional transmutation in the massless limit,  and it also possesses an infra-red conformal fixed point.

 In quantum field theory it is well known that in some cases, different regularization methods give different results for the same problem \cite{Peskin,Oliv13}. This is also the case for non-local quantum field theories \cite{Klep92,Oxman96,Martin95}. In \cite{Phi98}, it was shown that for the non-relativistic $\delta$- function potential in odd $d>3$, the dimensional regularization gives different results than the cutoff regularization, provided that the calculation is done in a non-perturbative setting.

In the case of the relativistic Schr\"odinger equation,  the Hamiltonian is nonlocal unlike the non-relativistic case. A naive guess would tell us that for this case, and for a certain problem, different regularization  methods  give different results. However, in this paper we prove that this is not the case for the solution of the relativistic  Schr\"odinger equation for a  point particle in an external 1-dimensional $\delta$-function potential. We study the solution in a non-perturbative setting using cutoff regularization. It gives exactly the same results from our previous calculations using dimensional regularization. This important because it is an evidence of universality.

The rest of this paper is organized as the following: Section 2 is for studying the
bound state problem non-relativistically and relativistically using contour integrals.
Section 3 is for discussing the gap equation and cutoff regularization, and using the bound state energy to define a renormalization
condition. Section 4 for studying the
scattering state problem non-relativistically and relativistically with the use of contour integrals, and defining the energy-dependence of the running coupling constant.
Section 5 is for the conclusions.

\section{The Bound State}
The non-relativistic solution of the Schr\"odinger equation for the $\delta$-function potential is a textbook problem. Nevertheless, we discuss it briefly to highlight the similarities and differences between the non-relativistic and relativistic case, especially regarding the contour integration approach used to calculate the wave function in both cases. The Schr\"odinger equation for this case is
\begin{equation}\label{SchrXNon}
\frac{p^2}{2m}\Psi (x)+\lambda \delta(x) \Psi (x)
=E\Psi(x),
\end{equation}
in momentum space, the above equation takes the following form
\begin{equation}\label{SchrPNon}
\frac{p^2}{2m}\widetilde{\psi }(p)+\lambda \psi (0)
=E\widetilde{\psi }(p),
\end{equation}
where
\begin{equation}\label{Fourier}
\Psi(x) = \frac{1}{2 \pi} \int dp \ \widetilde \Psi(p) \exp(i p x), \quad
\Psi(0) = \frac{1}{2 \pi} \int dp \ \widetilde \Psi(p). \
\end{equation}
Accordingly, in coordinate space the wave function for the bound state  is
\begin{equation}\label{PsideltaDerWF}
\Psi_B (x)=\frac{m\lambda }{\pi } \int_{-\infty }^{\infty }%
\frac{\Psi_B(0)e^{ipx}}{2m\Delta E_B-p^2}dp,
\end{equation}
where $\Delta E_B < 0$ is the binding energy,  $\Delta E_B =E_B-m \ll m $, and $E_B$ is the bound state energy. The value of $\Psi _B(x)$  in the above equation can be obtained using contour integral. The contour in this case has one pole inside the upper half circle at $p=i\sqrt{-2m\Delta E_B}$,  as it is illustrated in Figure.1 top panel. For the non-relativistic $\delta$-function potential, the wave function of the bound state is finite at the origin, and it is given by the following equation
\begin{equation}
\label{nonrelbound}
\Psi_B(x) = \sqrt{\varkappa} \exp(- \varkappa |x|), \quad
\Delta E_B = - \frac{\varkappa^2}{2m}=-\frac{m\lambda^2}{2},
\end{equation}
\begin{figure}[H]
\begin{center}
\epsfig{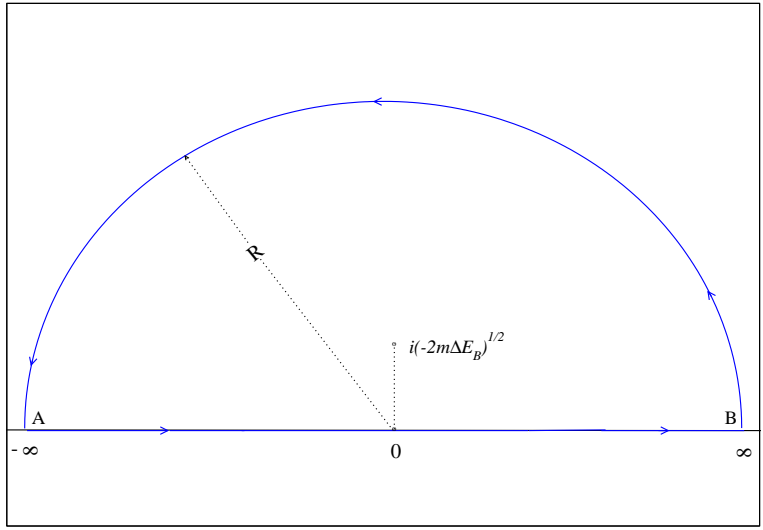} \vskip1.6 cm
\epsfig{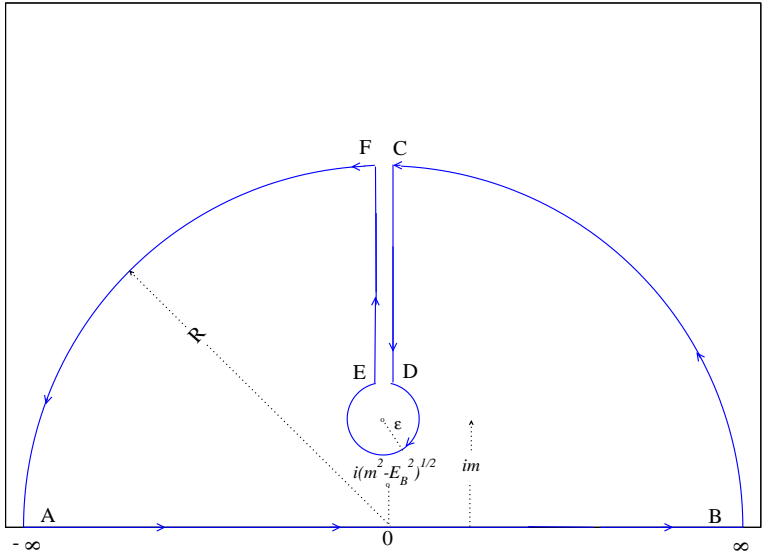}
\end{center}
\caption{\it The integration contours for obtaining the wave function of
the bound state. In the non-relativistic case, there is a pole inside the contour at $i\sqrt{-2m\Delta E_B}$, but no branch cut(top panel).
For relativistic case, there is a branch cut along the positive imaginary axis, starting at $p = im$, and there is a pole at $p = i \sqrt{m^2 - E_B^2}$ (bottom panel).}
\label{isofig2}
\end{figure}

The relativistic time-independent Schr\"odinger equation is
\begin{equation}
\sqrt{p^2 + m^2} \Psi(x) + \lambda \delta(x) \Psi(x) = E \Psi(x),
\end{equation}
where $\lambda$ is the bare coupling constant. In momentum space, the above equation is
\begin{equation}\label{RelSchrP}
\sqrt{p^2 + m^2} \tilde{\Psi}(p) + \lambda \Psi(0) = E \tilde{\Psi}(p).
\end{equation}
Solving for $\tilde{\Psi}(p)$, we get
\begin{equation}\label{Psip}
\widetilde \Psi_B(p) =
\frac{\lambda \Psi_B(0)}{E_B - \sqrt{p^2 + m^2}}.
\end{equation}
From the above equation, we express the bound state wave function in coordinate space as
\begin{equation}
\Psi_B(x) = \frac{\lambda \Psi_B(0)}{2 \pi} \int dp \
\frac{\exp(i p x)}{E_B - \sqrt{p^2 + m^2}}.
\end{equation}
Again here, the value of $\Psi_B (x)$  in the above equation can be obtained using contour integral.  For a bound state $0 < E_B < m$, the contour in this case has one pole inside the upper half circle at $p = i \sqrt{m^2 - E_B^2}$. In addition, it has a branch
cut along the positive imaginary axis starting at $p = i m$ as it is illustrated in Figure.1 (bottom panel). For this case $\Psi_B(x)$ is
\begin{equation}
\label{boundstate}
\Psi_B(x) = \lambda \Psi_B(0) \left[\frac{1}{\pi}  \int_m^\infty d\mu
\frac{\sqrt{\mu^2 - m^2}}{E_B^2 - m^2 + \mu^2} \exp(- \mu |x|) +
\frac{E_B \exp(- \sqrt{m^2 - E_B^2} |x|)}{\sqrt{m^2 - E_B^2}}\right].
\end{equation}
For a strong bound state when $-m<E_B<0$, and for an ultra-strong bound state when $-m>E_B$, the pole does not have a residue. For this case $\Psi_B (x)$ is
\begin{equation}
\label{boundstateH}
\Psi_B(x) =  \frac{\lambda \Psi_B(0)}{\pi}  \int_m^\infty d\mu
\frac{\sqrt{\mu^2 - m^2}}{E_B^2 - m^2 + \mu^2} \exp(- \mu |x|).
\end{equation}
Both wave functions are logarithmically divergent at the origin.

In the non-relativistic limit, $E_B-m\ll m=\Delta E_B=-\varkappa^2/2m$, the relativistic wave function of
eq.(\ref{boundstate}) reduces to
\begin{equation}
\Psi_B(x) = \sqrt{\varkappa}
\left[\frac{\varkappa}{m \pi}  \int_m^\infty d\mu
\frac{\sqrt{\mu^2 - m^2}}{\mu^2 - \varkappa^2} \exp(- \mu |x|) +
\exp(- \varkappa |x|)\right],
\end{equation}
where $\varkappa\ll m$.  The wave function indeed
reduces to the non-relativistic wave function of eq.(\ref{nonrelbound}).
However, the divergence of the relativistic wave function persists for any
non-zero value of $\varkappa/m$. On the other hand, the strong bound and the ultra strong bound states do not have a non-relativistic limit. They are purely relativistic states.
\section{The Gap Equation and Momentum Cutoff Regularization}
From eq.(\ref{Psip}) and eq.(\ref{Fourier}), we get
\begin{equation}
    \Psi_B(0) = \frac{1}{2 \pi} \int dp \ \widetilde \Psi_B(p) =
 \frac{\lambda \Psi_B(0)}{2 \pi} \int dp \ \frac{1}{E_B - \sqrt{p^2 + m^2}}.
\end{equation}
The above equation gives
\begin{equation}\label{Gap}
    \frac{1}{\lambda} =
\frac{1}{2 \pi} \int dp \ \frac{1}{E_B - \sqrt{p^2 + m^2}}=I(E_B),
\end{equation}
where $I(E_B)$ depends on the value of $E_B$. The above equation determines the relation between $\lambda$ and the bound state energy $E_B$. The integral in eq.(\ref{Gap}) is logarithmically ultra-violet divergent, therefore it needs to
be regularized. We do this by using the  momentum cutoff regularization. To proceed with this, first we must isolate the term that is causing the divergency. We do that by expanding the integrand in  eq.(\ref{Gap}) in powers of $E_B/\sqrt{p^2 + m^2}$, and then we get
\begin{equation}\label{IEBD}
I(E_B) = - \frac{1}{2\pi} \int_{-\Lambda} ^{\Lambda} \left(
\frac{1}{\sqrt{p^2 + m^2}}+
\sum_{n=0}^\infty \left(\frac{E_B}{\sqrt{p^2 + m^2}}\right)^n\right)dp,
\end{equation}
where $\Lambda$ is the momentum cutoff . By integrating term by term, and then taking the limit $\Lambda\rightarrow \infty$, we find that all terms under the summation sign are finite, while the first term is logarithmically ultra-violet divergent. The resultant series from the second term converges for $|E_B|<m$, and can be summed. Accordingly, we can write $I(E_B)$ as
\begin{equation}
\label{IEB}
I(E_B) = \frac{1}{2\pi }
\log\left(\frac{\sqrt{\Lambda^2+m^2}-\Lambda}{\sqrt{\Lambda^2+m^2}+\Lambda}\right)- \frac{E_B}{2 \pi \sqrt{m^2 - E_B^2}}
\left(\pi + 2 \arcsin\frac{E_B}{m}\right),
\end{equation}
or
\begin{equation}
\label{IEBG}
I(E_B) = \frac{1}{2\pi }
\log\left(\frac{\sqrt{\Lambda^2+m^2}-\Lambda}{\sqrt{\Lambda^2+m^2}+\Lambda}\right)+I_c(E_B),
\end{equation}
where $I_{c}(E_B)$ is the finite part of $I(E_B)$. In this case,
\begin{equation}
\label{IEBc0}
I_{c}(E_B) = -
 \frac{E_B}{2 \pi \sqrt{m^2 - E_B^2}}
\left(\pi + 2 \arcsin\frac{E_B}{m}\right)\hspace{6mm} |E_B|<m.
\end{equation}
For an ultra-strong bound state with energy $E_B < - m$ the series  diverges. Still, the result
can be obtained by directly integrating the convergent expression, and taking the limit $\Lambda\rightarrow \infty$
\begin{eqnarray}
\label{IEBU}
I_{c}(E_B)&=&\frac{1}{2 \pi} \int dp \ \left(\frac{1}{E_B - \sqrt{p^2 + m^2}} +
\frac{1}{\sqrt{p^2 + m^2}} \right) = \nonumber \\
&&\frac{E_B}{\pi \sqrt{E_B^2 - m^2}} \
\text{arctanh} \frac{\sqrt{E_B^2 - m^2}}{E_B}, \hspace{6mm} E_B<-m.
\end{eqnarray}
As a renormalization condition, we fix  the binding energy $E_B$  in
units of the mass $m$, such that the running bare coupling is given by
\begin{equation}
\frac{1}{\lambda(\Lambda)} = \log\left(\frac{\sqrt{\Lambda^2+m^2}-\Lambda}{\sqrt{\Lambda^2+m^2}+\Lambda}\right)- \frac{E_B}{2 \pi \sqrt{m^2 - E_B^2}}
\left(\pi + 2 \arcsin\frac{E_B}{m}\right).
\end{equation}
At the non-relativistic limit, the binding energy
$\Delta E_B = E_B - m$ is small compared to the rest mass. In this case, the
running bare coupling is given by \cite{Our}
\begin{eqnarray}
\frac{1}{\lambda(\Lambda)}&=&\frac{1}{2\pi } \log\left(\frac{\sqrt{\Lambda^2+m^2}-\Lambda}{\sqrt{\Lambda^2+m^2}+\Lambda}\right)
 - \frac{E_B}{2 \pi \sqrt{m^2 - E_B^2}}
\left(\pi + 2 \arcsin\frac{E_B}{m}\right) \nonumber \\
&\rightarrow&\frac{1}{2\pi} \log\left(\frac{\sqrt{\Lambda^2+m^2}-\Lambda}{\sqrt{\Lambda^2+m^2}+\Lambda}\right) -
\sqrt{\frac{m}{- 2 \Delta E_B}}.
\end{eqnarray}
Here it is very useful to compare the above equation with the non-relativistic contact interaction $\lambda \delta(x)$,
which does not require renormalization, for the attractive delta $\lambda < 0$, the bound state
energy is given by $\Delta E_B = - m \lambda^2/2$, or
$1/\lambda = - \sqrt{- m/2 \Delta E_B}$. This suggests to define a
renormalized coupling $\lambda(E_B)$ as
\begin{equation}
\label{lambdaEB}
\frac{1}{\lambda(E_B)} = - \frac{E_B}{2 \pi \sqrt{m^2 - E_B^2}}
\left(\pi + 2 \arcsin\frac{E_B}{m}\right) < 0,
\end{equation}
\section{The Scattering States}
For the non-relativistic scattering states, we assume that the solution in momentum space has the following ansatz \cite{Widmer}
\begin{equation}\label{GeneralAnsatzNon}
  \tilde{\Psi}(p) =\delta(p-\sqrt{2m\Delta E})+\delta(p+\sqrt{2m\Delta E})+\tilde{\Phi}_E(p),
\end{equation}
where $\Delta E=E-m\ll m$. After substituting eq.(\ref{GeneralAnsatzNon}) into eq.(\ref{SchrPNon}), we get
\begin{equation}
    \tilde{\Phi}_E(p) =
\frac{2 m\lambda}{\pi} \frac{1 + \pi \Phi_E(0)}{2m\Delta E - p^2}
\end{equation}
\begin{equation}\label{PsideltaDerWF}
\Phi_E (x)=\frac{2 m\lambda}{\pi} (1 + \pi \Phi_E(0))\int_{-\infty }^{\infty }%
\frac{e^{ipx}}{2m\Delta E-p^2}dp
\end{equation}
In the above equation, the value of $\Phi_E (x)$  can be obtained using contour integral. In this case, the contour has two poles on the real axis at $p=\pm\sqrt{2m\Delta E}$ as it is illustrated in Figure.2 top panel. After calculating $\Phi_E (x)$, it is straightforward to calculate $\Psi_E (x)$, which is
\begin{equation}\label{PsiScatNon}
\Psi_E(x) = A(k) \left[\cos(k x) + \frac{\lambda m}{k} \sin(k |x|)\right].
\end{equation}
For the non-relativistic $\delta$-function potential, the wave function of the scattering state is finite at the origin.
\begin{figure}[h]
\begin{center}
\epsfig{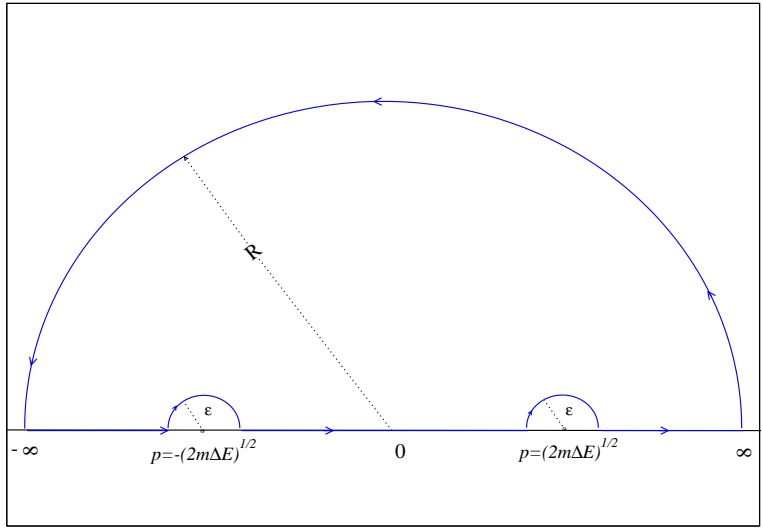} \vskip1.6 cm
\epsfig{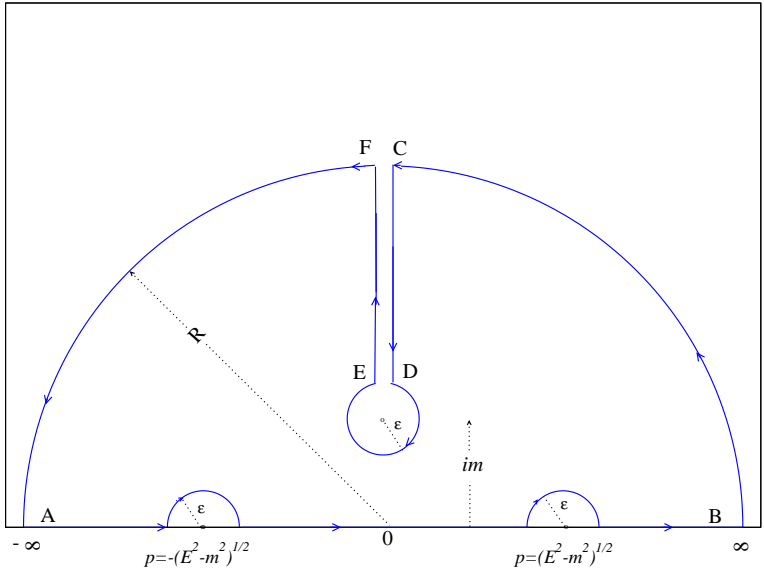}
\end{center}
\caption{\it The integration contours for obtaining the wave function of
the scattering states with $E>m$. In the non-relativistic case, there are two poles on the real axis at $\pm \sqrt{2m\Delta E}$, but no branch cut (top panel).
For relativistic case, there is a branch cut along the positive imaginary axis, starting at $p = im$, and there are two  poles on the real axis at $p =\pm  \sqrt{E^2 - m^2}$ (bottom panel).}
\label{isofig2}
\end{figure}
For the relativistic case, the ansatz takes the following form \cite{Our}
\begin{equation}\label{GeneralAnsatz}
  \tilde{\Psi}_E(p) =\delta(p-\sqrt{E^{2}-m^{2}})+\delta(p+\sqrt{E^{2}-m^{2}})+\tilde{\Phi}_E(p),
\end{equation}
by substituting eq.(\ref{GeneralAnsatz}) into eq.(\ref{RelSchrP}), and after solving for $\tilde{\Phi}_E(p)$, we get
\begin{equation}\label{PhiRel}
   \widetilde \Phi_E(p) =
\frac{\lambda}{\pi} \frac{1 + \pi \Phi_E(0)}{E - \sqrt{p^2 + m^2}},
\end{equation}
where $E>m$. From eq.(\ref{PhiRel}), and by using momentum cutoff regularization, we can write
\begin{equation}\label{PhizeroDelta}
    \Phi_E(0)=\frac{1}{2\pi}\int dp\tilde{\Phi}_E (p)= \frac{\lambda}{\pi}(1 + \pi \Phi_E(0))\int^{\Lambda}_{-\Lambda}\frac{dp}{E-\sqrt{p^2+m^2}}.
\end{equation}
The integral in the above equation is similar to the one in eq.(\ref{Gap}), with the exception that $E>0$. Therefore, we can write
\begin{equation}\label{IEE}
  I(E)= \frac{1}{2\pi} \int^{\Lambda}_{-\Lambda}\frac{dp}{E-\sqrt{p^2+m^2}}.
\end{equation}
As in the case of $I(E_B)$, the term causing the divergency is $1/\sqrt{p^2 + m^2}$. Therefore, we can write
\begin{equation}\label{IE0}
I(E) = \frac{1}{2\pi }
\log\left(\frac{\sqrt{\Lambda^2+m^2}-\Lambda}{\sqrt{\Lambda^2+m^2}+\Lambda}\right)+I_{c}(E),
\end{equation}
where $I_{c}(E)$ is the convergent part of $I(E)$ for $\Lambda\rightarrow\infty$. It can be obtained from eq.(\ref{IEBU}) after replacing $E_B$ by $E$, and we get
\begin{equation}\label{ICE}
   I_{c}(E)= \frac{E}{\pi \sqrt{E^2 - m^2}} \
\text{arctanh} \frac{\sqrt{E^2 - m^2}}{E}, \hspace{6mm} E>m.
\end{equation}
Now, we substitute for $\Phi_E(0)$ from eq.(\ref{PhizeroDelta}) into eq.(\ref{PhiRel}), and also substituting for $\lambda$  from eq.(\ref{Gap}), we get
\begin{equation}\label{PhiPdelta1final}
    \tilde{\Phi}_E (p)=\frac{1}{\pi}\frac{1}{I(E_B)-I(E)}\frac{1}{E-\sqrt{p^2+m^2}}.
\end{equation}
From eq.(\ref{IEBG}) and eq.(\ref{IE0}), it is obvious that  ultra-violet divergences
of $I(E)$ and $I(E_B)$ is canceled. Therefore, we can say in this case that the problem is renormalizable. Accordingly, eq.(\ref{PhiPdelta1final}) can be written as
\begin{equation}\label{PhiPdelta1final2}
    \tilde{\Phi}_E (p)=\frac{1}{\pi}\frac{1}{I_c(E_B)-I_c(E)}\frac{1}{E-\sqrt{p^2+m^2}},
\end{equation}
certainly $I_c(E_B)-I_c(E)$ is a finite quantity.
The value of $\Phi_E (x)$  can be obtained from eq.(\ref{PhiPdelta1final2}) using contour integral. The contour in this case has two poles on the real axis at $p = \pm \sqrt{E^2 - m^2}$, and it has  a branch cut along the positive imaginary axis starting at $p = i m$ as it is illustrated in Figure.2 bottom panel. After calculating $\Phi _E(x)$, it is straightforward to calculate $\Psi_E (x)$, which is
\begin{eqnarray}\label{PsiScat}
\Psi_E(x)&=&A(k) \left[\cos(k x) +
\frac{1}{I_c(E_B)-I_c(E)} \frac{\sqrt{k^2 + m^2}}{k} \sin(k |x|) \right.
\nonumber \\
&-&\left. \frac{1}{\pi(I_c(E_B)-I_c(E))} \int_m^\infty d\mu
\frac{\sqrt{\mu^2 - m^2}}{\mu^2 + k^2} \exp(- \mu |x|)\right], \
E = \sqrt{k^2 + m^2}.\nonumber\\
\end{eqnarray}
Again, the wave function is logarithmically divergent at the origin.
To understand the meaning of the expression $I_c(E_B)-I_c(E)$, we take the non-relativistic limit when $\Delta E=E-m\ll m$, and $|\Delta E_B|=E_B-m\ll m$, then
\begin{equation}\label{I-IEQ}
    \frac{1}{I_c(E_B)-I_c(E)}=-\sqrt{-\frac{2\Delta E_B}{m}}=\lambda,
\end{equation}
and the non-relativistic limit of eq.(\ref{PsiScat}) is
\begin{eqnarray}\label{PsiScatLimit}
\Psi_E(x)&=&A(k) \left[\cos(k x) +
\lambda \frac{m}{k} \sin(k |x|) \right.
\nonumber \\
&-&\left. \frac{\lambda}{\pi} \int_m^\infty d\mu
\frac{\sqrt{\mu^2 - m^2}}{\mu^2 + k^2} \exp(- \mu |x|)\right], \
\Delta E = \frac{k^2}{2m}.
\end{eqnarray}
in the non-relativistic limit, $\Psi_E(x)$ reduces to the non-relativistic wave function of eq.(\ref{PsiScatNon}).
However, the divergence of the relativistic wave function persists. Nevertheless, comparing eq.(\ref{PsiScatLimit}) with eq.(\ref{PsiScatNon}) together with eq.(\ref{I-IEQ}) is telling us that $(I_c(E_B)-I_c(E))^{-1}$ is nothing but the energy-dependent running coupling constant $\lambda(E,E_B)$ which is renormalized at the scale $E_B$, and finite for $\Lambda\rightarrow \infty$. Therefore, we have
\begin{eqnarray}
\label{lambdarun}
\lambda(E,E_B) = \frac{1}{I_c(E_B) - I_c(E)}&=&
- \left[\frac{E}{\pi \sqrt{E^2 - m^2}}
\text{arctanh}\frac{\sqrt{E^2 - m^2}}{E} \right. \nonumber \\
&+&\left. \frac{E_B}{2 \pi \sqrt{m^2 - E_B^2}}
\left(\pi + 2 \arcsin\frac{E_B}{m}\right)\right]^{-1}.
\end{eqnarray}
\section{Conclusions}
We proved that cutoff regularization gives the same results as dimensional regularization for the problem of the Schr\"odinger equation for a relativistic point particle in an
external 1-dimensional $\delta$-function potential. In both cases, the  ultra-violet divergences
have been canceled, and in both cases we get the same energy-dependent running coupling constant $\lambda(E,E_B)$, which is renormalized at the scale $E_B$, and finite. This means that the features that was obtained using dimensional regularization in \cite{Our} are preserved when using cutoff regularization. This can be considered as an evidence of universality for a nonlocal Hamiltonian.
\section*{Acknowledgments}
This publication was made possible by the NPRP grant \# NPRP 5 - 261-1-054 from
the Qatar National Research Fund (a member of the Qatar Foundation). The
statements made herein are solely the responsibility of the authors.

\end{document}